\documentclass[aps,prl,preprint,superscriptaddress]{revtex4}
\usepackage{amsmath,graphicx,textcomp}
\usepackage{placeins}
\usepackage{bbm}
\usepackage{xcolor}
\usepackage{float}
\begin{document}
	\title{Zeeman effect induced 0-$\pi$ transitions in ballistic Dirac semimetal Josephson junctions}
	
	\author{Chuan Li}
	\affiliation{MESA$^+$ Institute for Nanotechnology, University of Twente, The Netherlands}
	\author{Bob de Ronde}
	\affiliation{MESA$^+$ Institute for Nanotechnology, University of Twente, The Netherlands}
	\author{Jorrit de Boer}
	\affiliation{MESA$^+$ Institute for Nanotechnology, University of Twente, The Netherlands}
	\author{Joost Ridderbos}
	\affiliation{MESA$^+$ Institute for Nanotechnology, University of Twente, The Netherlands}
	\author{Floris Zwanenburg}
	\affiliation{MESA$^+$ Institute for Nanotechnology, University of Twente, The Netherlands}
	\author{Yingkai Huang}
	\affiliation{Van der Waals - Zeeman Institute, IoP, University of Amsterdam, The Netherlands}
	\author{Alexander Golubov}
	\affiliation{MESA$^+$ Institute for Nanotechnology, University of Twente, The Netherlands}
	\affiliation{Moscow Institute of Physics and Technology, Dolgoprudny, Moscow Region, 14170, Russia}
	\author{Alexander Brinkman}
	\affiliation{MESA$^+$ Institute for Nanotechnology, University of Twente, The Netherlands}
	
	\date{\today}
	
	\begin{abstract}
		One of the consequences of Cooper pairs having a finite momentum in the interlayer of a Josephson junction, is $\pi$-junction behavior. The finite momentum can either be due to an exchange field in ferromagnetic Josephson junctions, or due to the Zeeman effect. Here, we report the observation of Zeeman effect induced 0-$\pi$ transitions in Bi$_{1-x}$Sb$_x$, three-dimensional Dirac semimetal-based Josephson junctions. The large in-plane g-factor allows tuning of the Josephson junctions from 0- to $\pi$- regimes. This is revealed by measuring a $\pi $ phase shift in the current-phase relation  measured with an asymmetric superconducting quantum interference device (SQUID). Additionally, we directly measure a non-sinusoidal current-phase relation in the asymmetric SQUID, consistent with models for ballistic Josephson transport.
	\end{abstract}
	
	\maketitle
	The ground state of a Josephson junction can have a phase of $\pi $ for a large variety of physical reasons, such as the presence magnetic impurities \cite{Bulaevskii}, magnetic exchange fields in superconductor-ferromagnet-superconductor Josephson junctions \cite{FF,LO,Demler,Ryazanov}, occupation of orbital levels \cite{Delagrange}, unconventional order parameter symmetries, such as in corner junctions employing the $d$-wave order parameter symmetry in high-$T_c$ cuprates \cite{Harlingen}, and non-equilibrium population of quasiparticles \cite{Baselmans}. Only recently, it has been realized that the Zeeman effect can also be responsible for the 0-$\pi$ transition. Experimental indications of Zeeman effect-induced $\pi$-junctions  have been reported for Josephson junctions based on the topological insulator HgTe \cite{Yacoby}, Bi nanowires \cite{Murani} and the Dirac semimetal Bi$_{1-x}$Sb$_x$ \cite{Li2018}. 
	
	In principle, Zeeman effect causes the non spin-degenerate band structure to shift both in energy and in momentum. This momentum shift results in a dephasing of the superconducting pairing potential. As the dephasing increases, the sign of the pair potential changes, thus altering the ground state by a phase shift of $\pi$. In a Dirac semimetal, electrons follow a linear dispersion relation, i.e. $E_F=\hbar v_F k_F$. Upon applying a magnetic field, the Dirac cones shift in $k$-space and Cooper pairs acquire a finite momentum of $\Delta k=\pm\dfrac{g\mu_B B}{\hbar v_F}$, depending on their spin. When the phase $\Delta k l$ increases (where $l$ is the length of transport path) and the superconducting pair potential changes sign, the sign of the supercurrent in a Josephson junction changes. However, the superconducting coherence diminishes in magnetic field due to many other reasons (e.g. the orbital effect, scattering etc.), so to observe this transition a large g-factor is indispensable. The Dirac semimetal Bi$_{0.97}$Sb$_{0.03}$ gains a unique advantage by inheriting an extremely large g-factor along the bisectrix from bismuth \cite{Rowell1964,Roth1959,Zhu2014}. This can also be understood in terms of Andreev bound states, which are shifted in energy by the Zeeman effect. When the energy shift is comparable to the Andreev level spacing (usually characterized by Thouless energy $E_{Th}=(\hbar v_F)/L$), a complete anti-crossing of the Andreev levels can occur, resulting in a $0-\pi$ transition \cite{Kuplevachskii,Yokoyama}.
	
	Whereas our previous work showed oscillations in the supercurrent as $\Delta k $ increases with the magnetic field, the associated sign changes between the 0 and $\pi$ states have not yet been revealed.
	Here, we report the observation of Zeeman effect induced $0-\pi$ transitions by incorporating a Dirac semimetal-based junction in an asymmetric superconducting quantum interference device (SQUID), by which we can measure the current-phase relation (CPR) of the junction directly. The in-plane magnetic field can shift the phase of the SQUID oscillations (of the out-of-plane magnetic field) by $\pi$. Moreover, the CPR was measured to be non-sinusoidal, quantitatively consistent with models for ballistic Josephson junctions.
	
	The CPR in an SIS or SNS junction in different limits has been studied intensively. In the tunneling regime, the CPR is represented by $I_s = I_0\sin(\phi)$, where $I_0$ is the critical current amplitude of the sample junction and $\phi$ is the phase difference between the two superconducting electrodes. In an ideal situation, $\phi \equiv 2\pi\varPhi/\phi_0$ is determined by the magnetic flux that goes through the SQUID, $\varPhi$. A non-sinusoidal CPR was found in both the dirty \cite{KO-1} and the clean limit \cite{KO-2} of the Kulik-Omelyanchuk model (see also the review of \cite{Golubov2004}). With a perfect transparency, D = 1, the CPR of the point contact junction in the clean limit is just a segment of a sine function. The non-sinusoidal effect is strongest at high transparency and decreases as the interface becomes more opaque, until the CPR eventually becomes sinusoidal at very low transparency. Later, the theory was extended to the SNS junction, with a normal metal material as the interlayer. It was shown that, for a long SNS junction (i.e. $\xi_0 \ll d\ll\xi_T$, where $d$ is the distance between the superconducting leads, $\xi_0 = \hbar v_F/\Delta$ is the superconducting coherence length in the normal metal and $\xi_T = \xi_0T_c/T$ is the thermal coherence length), the CPR becomes linear within one period:
	$ I_s (\phi) = eN_2v_2\Delta_1\{\dfrac{\phi}{2}-\pi \, \text{Int}[\dfrac{\phi}{2\pi}+\dfrac{1}{2}]\}$
	\cite{Golubov2004,Ishii1970}. Thus, in clean SNS junctions, the CPR has a saw-tooth shape at low temperature. This can also be understood as higher order harmonic contributions ($\sin2\phi, \sin3\phi ...$) in the supercurrent. If disorder is included (diffusive/dirty limit), then the CPR will be smoother and closer to a sinusoidal shape.  The current amplitude also decreases rapidly with the increase of temperature and the saw-tooth shape changes gradually to a sinusoidal shape \cite{Bardeen1972}. 
	
	We fabricated a direct-current (dc) SQUID to measure the CPR in a Dirac semimetal-based Josephson junction. For this purpose, Bi$_{0.97}$Sb$_{0.03}$ single crystals were grown using a modified Bridgeman method (see Ref. \onlinecite{Li2018} for details). Thin flakes were exfoliated onto a Si substrate with a SiO$_2$ layer on top. Niobium (Nb) leads were fabricated by standard e-beam lithography and lift-off. The Dirac semimetal Josephson junction is incorporated in a Nb loop with an area of about 0.5 $\mu \textrm{m}^2$ (the configuration of the sample is illustrated in Fig.1b) As the second junction in the SQUID we made a Nb constriction that acts as a reference junction with a critical current of about $I_r =7 \mu A$, much larger than the critical current in Dirac semimetal flake, $I_c = 1.5 \mu A$. Here we define the amplitude of the supercurrent to be the critical current $I_c=$Max$[|I_s(\phi)|]$. When the SQUID is strongly asymmetric ($I_s\ll I_r$), the modulation in the critical current represents the CPR of the sample junction \cite{Della2007}.
	
\begin{figure}[h]
		\includegraphics[clip=true,width=15cm]{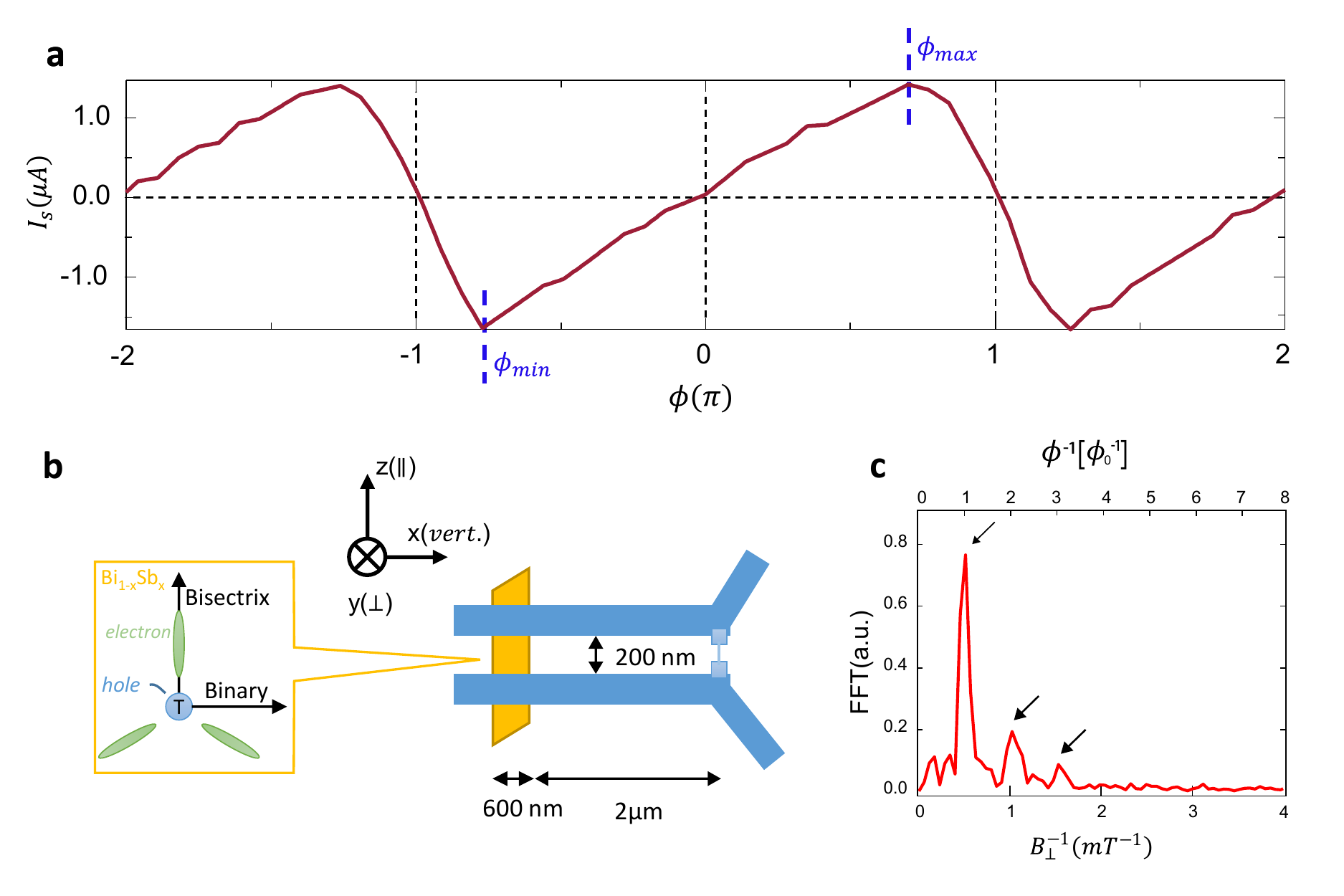}
		\caption{\textbf{Ballistic junction in SQUID configuration.} \textbf{(a)} The extracted critical current $I_s$ of the Bi$_{0.97}$Sb$_{0.03}$ junction as a function of the phase across the junction at 20 mK. The saw-tooth shape indicates the higher harmonics components in the supercurrent. The dashed blue lines indicate the maximum and the minimum of the critical current, which we used to estimate the skewness of the CPR shape in the main text.  \textbf{(b)} Scheme of the sample configuration. The niobium leads are normal to the Bi$_{0.97}$Sb$_{0.03}$ flake (orange) edges which is in parallel with a niobium constriction (light blue). x, y, z direction are assigned to vertical (vert.), perpendicular ($\perp$) and parallel ($\parallel$) direction respectively. The crystalline orientations in reciprocal space are indicated in the orange box, which has the same with coordinates as that in real space. The contacts are always aligned in such way that the current flow is along the bisectrix axis of the Bi$_{0.97}$Sb$_{0.03}$ crystals. \textbf{(c)} Fourier transform of the periodic $I_s(B)$, higher order harmonics are indicated by arrows. }
		\label{Fig:fig1_ballistic}
	\end{figure}

	In Fig.\ref{Fig:fig1_ballistic}a, we show the measured CPR when the parallel field $B_{\parallel} = 0$. The phase difference is presented in units of the flux quantum (see SM for the details). In an asymmetric SQUID, the total critical current does not drop to zero but oscillates around the $I_{r}$. To better capture the CPR of the sample junction, we subtract the reference current $I_{r}$, obtained by a strong smoothing. A saw-tooth shaped CPR is observed in our experiment. Correspondingly, the Fourier transform shows peaks at higher order harmonic frequencies (Fig.\ref{Fig:fig1_ballistic}c).
	
	In practice, the shape of the total critical current ($I_c^{tot} (\phi)$) as a function of the phase difference can be affected by the inductance of the superconducting loop. The inductance parameter is defined as $\beta_L = \dfrac{2\pi}{\phi_0}LI_c$. 
	At low temperatures, the inductance $L$ is dominated by the kinetic inductance ($L_k$) in superconducting devices. We estimate the $L_k$ in our experiment as $L_k = (\dfrac{m}{2n_se^2})(\dfrac{l}{A})\lesssim $1pH at 10mK, where $m$ is the Cooper pair mass, $n_s$ is the density of Cooper pair, $l$ is the length of the Nb wire and $A$ is the cross-section of the wire. In general, the superconducting loop in our experiment is very small. We then find $\beta_L \ll 1$, for which the deformation of the CPR by the inductance effect is negligible.
	
\begin{figure}[H]
		\includegraphics[clip=true,width=17cm]{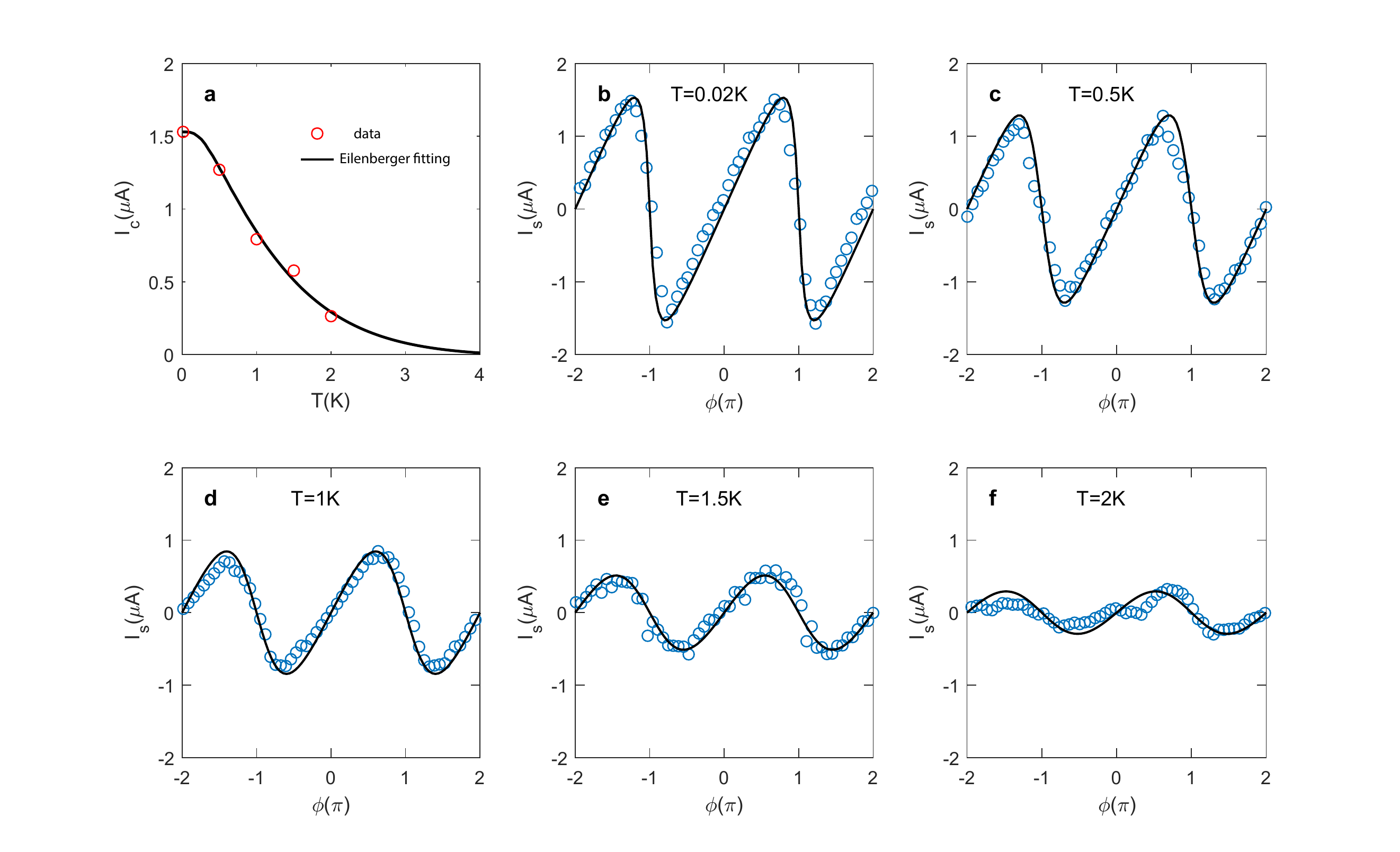}
		\caption{\textbf{Current-phase relation at different temperatures.} \textbf{(a)} Temperature dependence of the extracted critical current amplitude $I_c(T)\equiv$ Max$[|I_s(\phi)|]$. \textbf{(b)-(f)} Current-phase relation at different temperatures. Data (blue circles) is fitted using the same parameters extracted from \textbf{(a)}.}
		\label{Fig:fig2_CPRfit}
	\end{figure}

	To fit the data, we used a model \cite{Galaktionov} based on the Eilenberger equations for ballistic transport at arbitrary junction length and arbitrary interface transparencies. The results are shown in Fig.\ref{Fig:fig2_CPRfit}. We first fit the temperature dependence of the extracted critical current amplitude  $I_c(T)$ (Fig.\ref{Fig:fig2_CPRfit}a). A very high transparency $D\simeq 0.98$ and induced superconducting gap $\Delta \simeq 4.5K$ are obtained. Such a high transpancy is typical for this type of device. In the model, the induced superconducting gap near the contacts was used, together with the fact that the transmission between the proximitized and the normal Bi$_{0.97}$Sb$_{0.03}$ is naturally high. The coherence length $\xi_s \sim 100 \textrm{nm} \sim L$. Then, using the same parameters, we plot the simulated CPR data with the experimental data at different temperatures (Fig.\ref{Fig:fig2_CPRfit}b-f). We find a good and self-consistent agreement between the simulated and experimental results, not only for the amplitude of $I_s(\phi )$ but also for the shape, confirming ballistic transport in the junction.
	
\begin{figure}[H]
		\includegraphics[clip=true,width=15cm]{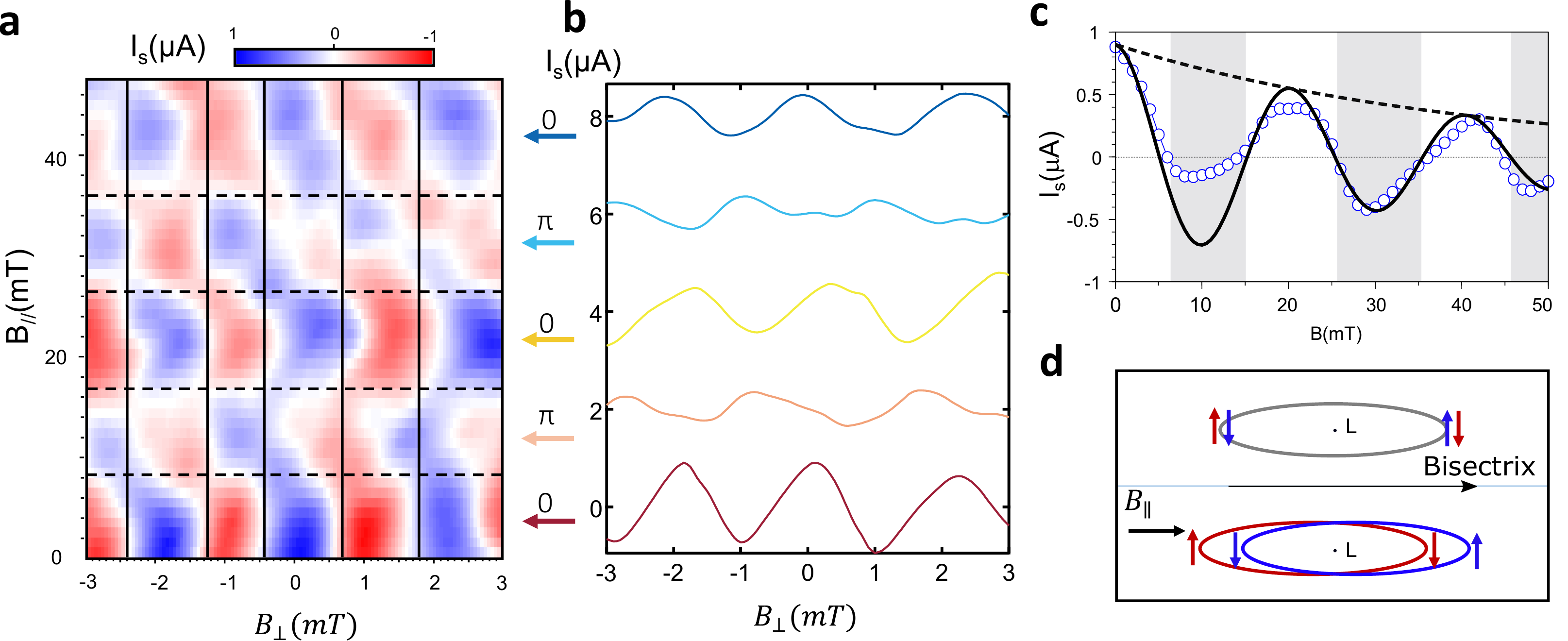}
		\caption{\textbf{Zeeman effect induced 0-$\pi$ transition.} \textbf{(a)} 2D color-plot of the Current-phase relation of Bi$_{0.97}$Sb$_{0.03}$ junction in a parallel magnetic field.  The $\pi$-shift in phase occurs periodically as the parallel field ($B_{\parallel}$) increases. The horizontal dashed lines separate the map into different 0 or $\pi$ regions.  \textbf{(b)} CPR curves at different parallel field values (indicated by the arrows). The junction alternates between 0 and $\pi$ states. For clarity, the data has been shifted vertically by successive increments of 2$\mu$A. \textbf{(c)} Extracted $I_s$ as a function of the parallel field $B_{\parallel}$. Blue circles: $I_s(B_{\parallel})$ at $B_{\perp}=0$; dashed line: fitting curve, exponential decay; solid line: fitting curve, decay and oscillation, taking into account of the hole-pocket channel contribution (see main text). \textbf{(d)} Schematic illustration of the Zeeman effect induced finite-momentum pairing. At zero field (upper), the Dirac Fermion is spin degenerate (electron pocket at the L-point); as an in-plane field is applied along the large g-factor direction (lower), the Dirac Fermi surface splits with polarized spin. Consequently, the formed Cooper pairs acquire a finite total momentum, driving the $0-\pi$ transition. }
		\label{Fig:fig3_Zeeman}
	\end{figure}

	Now we discuss the evolution of CPR in parallel magnetic field ($B_{\parallel}$, along the current flow direction in the junction). The results are shown in Fig.\ref{Fig:fig3_Zeeman}. In the 2D color plot, horizontal and vertical lines are drawn to indicate the nodes of the oscillations in parallel and perpendicular field respectively. In Fig.\ref{Fig:fig3_Zeeman}b, the CPRs for different parallel field values (indicated by arrows) are plotted. From these CPRs, it is apparent that at specific $B_{\perp}$ values, the CPR changes from a maximum to a minimum, suggesting a $\pi$-phase shift each time. 
	
	
	For bismuth, the g-factor of the electron pocket at the L-point is found to be highly anisotropic and extremely large along the bisectrix \cite{Roth1959,Zhu2014}. Since the Zeeman effect in Bi$_{0.97}$Sb$_{0.03}$ is very similar to that in pure Bi, we use a value typically found in literature ($ g\sim800-1000$ along the highly elongated direction). Because of this large g-factor, the Zeeman energy becomes relevant at very low magnetic fields.
	
	The critical current changes with $B$ as $I_s(B) \propto e^{-B/B_{eff}}$, where $B_{eff}^{-1} = B_1^{-1}+iB_2^{-1}$. Here, the two characteristic field scales are $B_1$, which corresponds to the amplitude decay of $I_c$, and $B_2$, which describes the period at which the junction alternates between the 0- and the $\pi$- states. In the clean limit ( $B_1 \gg B_2$), we expect no decay in the critical current amplitude, while in the diffusive regime ($B_1\sim B_2$), the oscillations are periodic in $\sqrt{B}$ \cite{Buzdin_SFS1992}. In the intermediate regime, the critical current can be described as a cosine oscillation with an exponential decay: 
	$ I_s(B_\parallel) \propto e^{-B/B_1}\cdot\cos{(B/B_2)}$. Fitting this expression for $I_s(B_\parallel)$ to the data, as in Fig.\ref{Fig:fig3_Zeeman}c, we obtain that $B_1 = 41 mT$ and $B_2 = 3.2 mT$. This indicates that the junction is almost in the clean limit.
	
\begin{figure}[H]
		\includegraphics[clip=true,width=15cm]{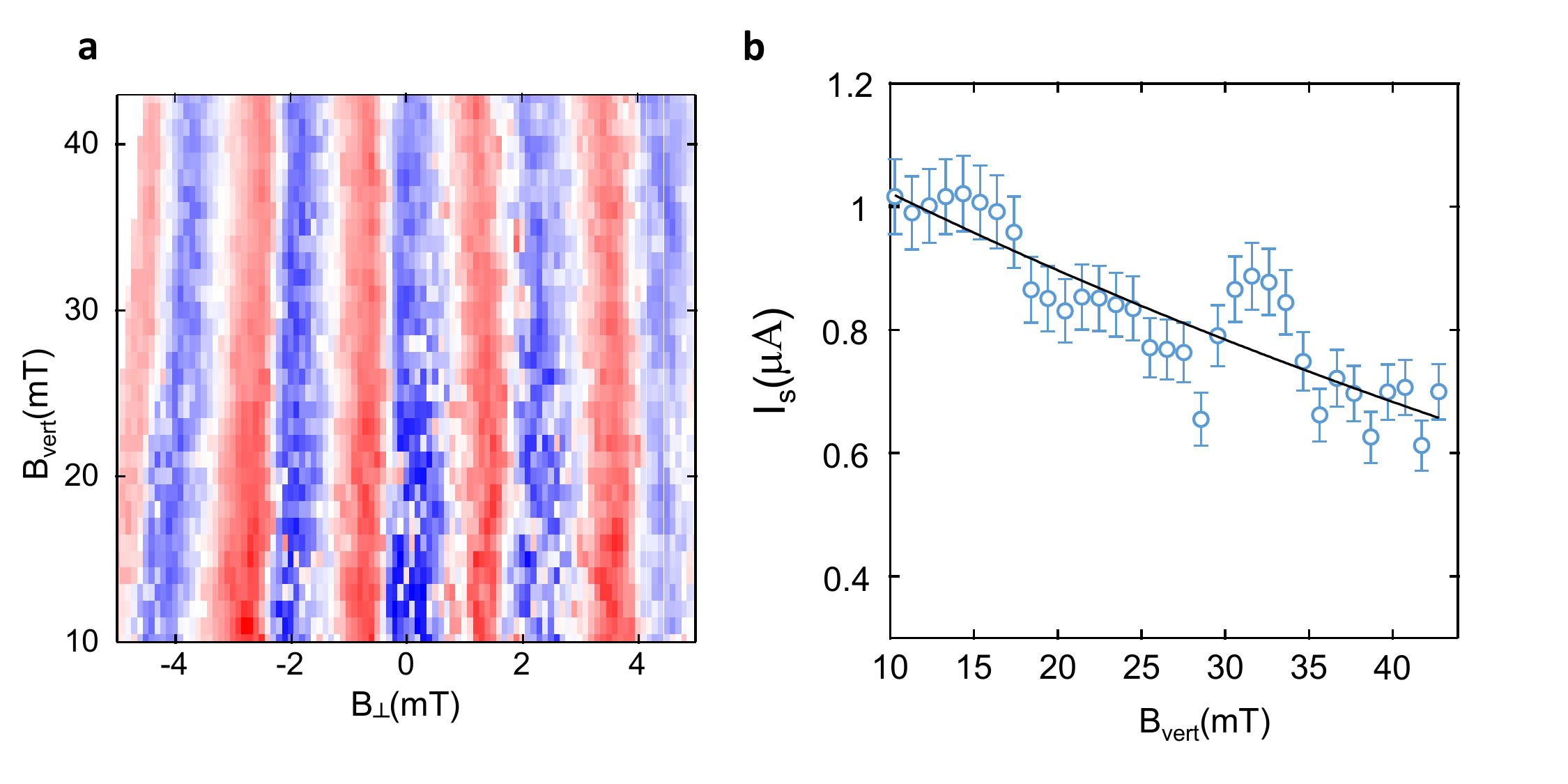}
		\caption{\textbf{Current-phase relation (CPR) in vertical field.} \textbf{(a)} 2D color-plot of the CPR in in-plane vertical field ($B_{vert}$). \textbf{(b)} Extracted $I_s(B_{vert})$ at $B_{\perp}=0$. $I_s$ decreases with increase of $B_{vert}$, but no change of sign in critical current $I_s$ is observed. Circles: experimental data, with error bars; black line: fitting curve.}
		\label{Fig:fig4_Bx}
	\end{figure}

	Due the high anisotropy of the g-factor, the response of the critical current should change dramatically upon applying a vertical in-plane field ($B_{vert}$). We show the measurement results in Fig.\ref{Fig:fig4_Bx}. The 2D color-plot of the CPR does not show the checkerboard pattern anymore, but instead features straight stripes which correspond to fixed phases ($\phi$). There is no $0-\pi$ transition within the same range of vertical field ($B_{vert}$) as parallel field ($B_{\parallel}$). This can be easily understood as a consequence of the small g-factor in that direction. In this case, $B_2$ becomes larger than $B_1$, and the field dependence of $I_c$ is dominated by the scattering characteristic field $B_1$. A reasonably good fit is obtained for $B_1 = 45mT$, which is similar to the parallel field value. This large $B_1$ field implies that the transport (and scattering) properties are homogeneous throughout the Bi$_{0.97}$Sb$_{0.03}$ junction. Note that we corrected the field misalignment, which can be noticed as slightly tilted stripes, with a correction angle of about 1 degree. The non-negligible shift near zero field is attributed to magnetic flux trapping, which shifts the ``zero field" to a random value (see Supp. mat. for details).
	
	Interestingly, we found that the critical current amplitude of the first $\pi$-state is smaller than the 0-state. And the shape of the CPR is gradually changing from the regular CPR as a function of magnetic field. This deformation of the CPR during the $0-\pi$ transition has been observed in superconducting quantum dots and is understood as a combination of the $0$- and $\pi$-states as the intermediate transition state \cite{Delagrandge2018,Maurand2012prx}. This coincides with the fact that part of the contribution to the critical current is from the hole pocket at the T-point in Bi$_{0.97}$Sb$_{0.03}$. The in-plane g-factor of the hole pocket is very small ($<2$)\cite{Rowell1964}. Therefore the Zeeman effect in this channel should be negligible and no $0-\pi$ transition should occur in our measured field range. That could explain why the critical current amplitude at the $\pi$-state (which originates from the electron pocket) dominates the total critical current and is slightly reduced at low fields in the $\pi$-state (The remaining 'trivial' 0-state current from the other pockets counteracts the $\pi$-state current). 
	
	It is worth noting that a system with spin-orbit coupling (SOC) can essentially break both time-reversal symmetry and inversion symmetry when an external magnetic field is applied. Consequently, the junction can alter not only between 0- and $\pi$-states, but can also have an additional phase shift, $\phi_0$ \cite{Buzdin2008}. Thus, a so-called $\phi_0$-junction can form. This effect was first observed in quantum dots (e.g. carbon nanotube, InAs nanowire-based devices). Recently, observations are also reported in Bi-based systems, e.g. Bi-nanowires \cite{Murani} and topological insulator Bi$_2$Se$_3$ \cite{Assouline2019}. In these systems with a reduced dimensionality, a strong Rashba-type spin-orbit coupling arises, which is lacking in the bulk system. Since in a Bi$_{0.97}$Sb$_{0.03}$ junction, the supercurrent is mainly carried by the bulk states, only the atomic SOC can play a role, which is at least two order of magnitude smaller than the reported Rashba effect \cite{Ishizaka2011}. Therefore, we would expect the SOC-induced phase shift in CPR to be small in our measurements.
	
	In summary, by incorporating a Nb-Bi$_{0.97}$Sb$_{0.03}$-Nb Dirac semimetal Josephson junction into an asymmetric SQUID, we have been able to measure the CPR and provide, for the first time, unambiguous evidence of the $0-\pi$ transitions associated with the finite momentum of the Cooper pairs due to the dominant Zeeman effect in parallel magnetic fields. Owing to the anisotropic g-factor of the Dirac cone at the crystallographic L-point, we can also explain the different effects of the two in-plane magnetic field orientations. The junction critical current as a function of temperature, as well as the CPR at all temperatures, were self-consistently fitted by models for ballistic transport. These results provide a clear physical picture of Bi$_{0.97}$Sb$_{0.03}$-based Josephson junctions, which supports phase-controled operations in topological Josephson devices in the future.
	
	\section{Acknowledgments}
	
	This work was financially supported by the Netherlands Organization for Scientific Research (NWO) through a VENI grant and the European Research Council (ERC) through a Consolidator Grant. We acknowledge the COST project “Nanoscale coherent hybrid devices for superconducting quantum technologies”—Action CA16218.
	We also thank S. Gueron and H. Bouchiat for helpful discussions and suggestions.


\begin{thebibliography}{1}
		\bibitem{Bulaevskii} L.N. Bulaevskii, V.V. Kuziǐ, A.A. Sobyanin, \emph{JETP Letters } \textbf{25} 290 (1977).
		\bibitem{FF} P. Fulde, R.A. Ferrell, \emph{Phys. Rev. }\textbf{135}, A550 (1964).
		\bibitem{LO} A.I. Larkin, Y.N. Ovchinnikov, \emph{Zh. Eksp. Teor. Fiz.} \textbf{47}, 1136 (1964).
		\bibitem{Demler} E.A. Demler, G.B. Arnold, M.R. Beasley, \emph{Phys. Rev. B} \textbf{55}, 15174 (1997).
		\bibitem{Ryazanov} V.V. Ryazanov, V.A. Oboznov, A.Yu. Rusanov, A.V. Veretennikov, A.A. Golubov, J. Aarts, \emph{Phys. Rev. Lett.} \textbf{86}, 2427 (2001).
		\bibitem{Delagrange} R. Delagrange, R. Weil, A. Kasumov, M. Ferrier, H. Bouchiat, and R. Deblock \emph{Phys. Rev. B} \textbf{93}, 195437 (2016).
		\bibitem{Harlingen} D.J. Van Harlingen, \emph{Rev. Mod. Phys.} \textbf{67}, 515 (1995).
		\bibitem{Baselmans} J.J.A. Baselmans, A.F. Morpurgo, B.J. van Wees, T.M. Klapwijk, \emph{Nature} \textbf{397}, 43 (1999).
		\bibitem{Yacoby} S. Hart \textit {et al.}, \emph{Nature Phys.} \textbf{13}, 87 (2017).
		\bibitem{Murani} A. Murani \textit {et al.}, \emph{Nature Communications} \textbf{8}, 15941 (2017).
		\bibitem{Assouline2019} Alexandre Assouline \textit{et al.}, \emph{Nature communications} \textbf{10}, 126 (2019).
		\bibitem{Li2018} Chuan Li et al. \emph{Nature Materials} \textbf{17}, 875 (2018).
		\bibitem{Kuplevachskii} Kuplevachskii and Falko \emph{JETP Lett.} \textbf{52}, 6 (1990).
		\bibitem{Yokoyama} Yokoyama \textit{et al.} \emph{Phys. Rev. B} \textbf{89}, 195407 (2014).
		\bibitem{KO-1} 	Kulik, I. O., and A. N. Omelyanchuk. Current flow in long superconducting junctions. \emph{Sov. Phys. JETP} \textbf{41}, 1071 (1975).
		\bibitem{KO-2} 	Kulik, I. O., and A. N. Omelyanchuk. Properties of superconducting microbridges in the pure limit. \emph{Sov. J. Low Temp. Phys. } \textbf{3}, 459 (1977).
		\bibitem{Golubov2004} 	A.A.Golubov \textit{et al.} The current-phase relation in Josephson junctions. \emph{Rev. of Mod. Phys. } \textbf{76}, 411 (2004).	
		\bibitem{Ishii1970}  Chikara Ishii. Josephson Currents through Junctions with Normal metal Barriers. \emph{Prog. Theor. Phys.} \textbf{44}, 1525 (1970).
		\bibitem{Bardeen1972} J. Bardeen and J.L.Johnson. Josephson current flow in pure superconducting-normal-superconducting junction. \emph{Phys. Rev. B} \textbf{5}, 72 (1972).
		\bibitem{Della2007} Della Rocca, M.L. \textit{et al.} Measurement of the current-phase relation of superconducting atomic contacts. \emph{Phys. Rev. Lett.} \textbf{99}, 127005 (2007).	
		\bibitem{Buzdin_SFS1992} A. Buzdin \textit{et al.}, \emph{Sov. Phys. JETP} \textbf{74}, 124 (1992).
		\bibitem{Delagrandge2018}  R.Delagrange \textit{et al.} $0-\pi$ Quantum transition in a carbon nanotube Josephson junction: Universal phase dependence and orbital degeneracy \emph{Physica B} \textbf{236}, 211 (2018).	
		\bibitem{Maurand2012prx}Romain Maurand \textit{et al.} First-Order 0-$\pi$ Quantum Phase Transition in the Kondo Regime of a Superconducting Carbon-Nanotube Quantum Dot \emph{Phys. Rev. X} \textbf{2}, 011009 (2012).	
		\bibitem{Rowell1964} G.E.Smith, G.A.Baraff and J.M.Rowell, Effective $g$ factor of electrons and holes in Bismuth. \emph{Phys. Rev.} \textbf{135}, 4A (1964).	
		\bibitem{Roth1959} L.M.Roth, B.Lax and S. Zwerdling, Theory of Optical Magneto-Absorption Effects in Semiconductors. \emph{Phys. Rev.} \textbf{114}, 90 (1959).	
		\bibitem{Zhu2014} Zengwei Zhu, \textit{et al.} Angle-resolved Landau spectrum of electrons and holes in bismuth. \emph{Phys. Rev. B} \textbf{84}, 115137 (2011).	
		\bibitem{Galaktionov} Artem V. Galaktionov and Andrei D.Zaikin, Quantum interference and supercurrent in multiple-barrier proximity structures. \emph{Phys. Rev. B} \textbf{65}, 184507 (2002).
		\bibitem{Buzdin2008} A. Buzdin, \emph{Phys. Rev. Lett.} \textbf{101}, 107005 (2008).
		\bibitem{Ishizaka2011} K. Ishizaka \textit{et al.}, \emph{Nature Materials} \textbf{10}, 521 (2011).
		
	\end{thebibliography}
\end{document}